\documentclass[11pt,english]{article}
\usepackage[T1]{fontenc}
\usepackage[latin1]{inputenc}
\usepackage{geometry}
\usepackage{setspace}
\makeatletter

\usepackage{babel}

\usepackage{babel}

\begin{document}

\title{Probabilistic Inequalities and Upper Probabilities in Quantum Mechanical
Entanglement}

\author{\textbf{J. Acacio de Barros}%
\thanks{E-mail: barros@sfsu.edu %
}\\
 Liberal Studies Program\\
 San Francisco State University\\
 San Francisco, CA 94132 \\
 \textbf{Patrick Suppes}%
\thanks{E-mail: psuppes@stanford.edu %
}\\
 CSLI - Ventura Hall\\
 Stanford University\\
 Stanford, CA, 94305-4115 USA}
\maketitle
\begin{abstract}
In this paper we analyze the existence of joint probabilities for
the Bell-type and GHZ entangled states. We then propose the usage
of nonmonotonic upper probabilities as a tool to derive consistent
joint upper probabilities for the contextual hidden variables. Finally,
we show that for the extreme example of no error, the GHZ state allows
for the definition of a joint upper probability that is consistent
with the strong correlations. 
\end{abstract}

\section{Introduction}

The issue of the completeness of quantum mechanics has been a subject
of intense research for almost a century. One of the most influential
papers is undoubtedly that of Einstein, Podolski and Rosen (1935),
where, after analyzing entangled two-particle states, concluded that
quantum mechanics could not be considered a complete theory. In 1963
John Bell showed that not only was quantum mechanics incomplete but,
if one wanted a complete description of local reality, one would obtain
correlations that are incompatible with the ones predicted by quantum
mechanics. This happens because some quantum mechanical states do
not allow for the existence of joint probability distributions of
all the possible outcomes of experiments. If a joint distribution
were to exist, then one could consistently create a local hidden variable
that would factor this distribution. The nonexistence of local hidden
variables that would {}``complete'' quantum mechanics, and hence
the nonexistence of joint probability distributions, was confirmed
experimentally by Aspect, Dalibard, and Roger (1982). They showed,
in a series of beautifully designed experiments, that an entangled
photon state of the form \[
|\psi\rangle=\frac{1}{\sqrt{2}}(|+-\rangle-|-+\rangle),\]
 (where $|+-\rangle\equiv|+\rangle_{A}\otimes|-\rangle_{B}$ represents,
for example, two photons $A$ and $B$ with helicity $+1$ and $-1,$
respectively) violates the Clauser-Horne-Shimony-Holt form of Bell's
inequalities, as predicted by quantum mechanical computations.

The nonexistence of joint probability distributions also comes into
play in the consistent-history interpretation of quantum mechanics.
In this interpretation, each sequence of properties for a given quantum
mechanical system represents a possible history for this system, and
a set of such histories is called a family of histories. A family
of consistent histories is one that has a joint probability distribution
for all possible histories in this family. One can easily show that
quantum mechanics implies the nonexistence of such probability functions
for some families of histories. Families of histories that do not
have a joint probability distribution are called inconsistent histories.

Another important example, also related to the nonexistence of a joint
probability distribution, is the famous Kochen-Specker theorem, which
shows that a given hidden-variable theory that is consistent with
the quantum mechanical results has to be contextual, i.e., the hidden
variable has to depend on the values of the actual experimental settings,
regardless of how far apart the actual components of the experiment
are located.

More recently, a marriage between Bell's inequalities and the Kochen-Specker
theorem led to the Greenberger-Horne-Zeilinger (GHZ) theorem. The
GHZ theorem shows that if one assumes that one can consistently assign
values to the outcomes of a measurement before the measure is performed,
a contradiction arises. Once again, having a complete data table would
allow us to compute the joint probability distribution, and therefore
no joint distribution exists that is consistent with quantum mechanical
results.

Although it is sometimes remarked that all the above contradictions
hold only for noncontextual hidden variable theories, a general proof
of this is not available. In this paper, we propose the usage of nonmonotonic
upper probabilities as a tool to derive consistent joint upper probabilities
for the contextual hidden variables.

\section{Upper Probabilities and Bell-type entanglement\label{sec:Upper-Bell-entanglement}}

We saw in the previous section that, for some cases, quantum mechanics
does not allow the existence of a joint probability distribution for
all the observables. However, if we weaken the probability axioms,
Suppes and Zanotti (1991) proved that a consistent set of upper probabilities
for the events can be found. Upper probabilities are defined in the
following way. Let $\Omega$ be a nonempty set, $F$ a boolean algebra
on $\Omega$, and $P^{*}$ a real valued function on $F.$ Then the
triple $(\Omega,F,P^{*})$ is an upper probability if for all $\xi_{1}$
and $\xi_{2}$ in $F$ we have that 
\begin{description}
\item [{(i)}] $0\leq P^{*}(\xi_{1})\leq1,$ 
\item [{(ii)}] $P^{*}(\emptyset)=0,$ 
\item [{(iii)}] $P^{*}(\Omega)=1,$ 
\end{description}
and if $\xi_{1}$ and $\xi_{2}$ are disjoint, i.e. $\xi_{1}\cap\xi_{2}=\emptyset,$
then 
\begin{description}
\item [{(iv)}] $P^{*}(\xi_{1}\cup\xi_{2})\leq P^{*}(\xi_{1})+P^{*}(\xi_{2}).$ 
\end{description}
As we see, property (iv) weakens the standard additivity axiom for
probability. Since monotonicity is one of the consequences of the
standard probability axioms, it may be true for an upper probability
that

\[
\xi_{1}\subseteq\xi_{2}\mbox{\ensuremath{\,}and\ensuremath{\,}}P^{*}(\xi_{1})>P^{*}(\xi_{2}).\]

Let us see how upper probabilities can be used to obtain joint upper-probability
distributions. We start with Bell's observables, represented by the
random variables $\mathbf{X},$ $\mathbf{Y}$ and $\mathbf{Z}$. Each
random variable corresponds to a different angle of measurement for
the Stern-Gerlach apparatus (we follow the example in Suppes and Zanotti
(1981)). Bell's thought experiment consisted of a two-particle system
with an entangled spin state. Since each random variable corresponds
to different spin orientations, we can only measure two of them at
the same time. Additionally, Bell's states are such that the expected
values of $\mathbf{X}$, $\mathbf{Y}$, and $\mathbf{Z}$ are zero,
and we have the constraint \begin{equation}
P(\mathbf{X}=1)=P(\mathbf{Y}=1)=P(\mathbf{Z}=1)=\frac{1}{2}.\label{eq:individual-marginals}\end{equation}

The question that Bell posed is whether we can fill the missing values
of the data table in a way that is consistent with the marginal distributions
given by quantum mechanics for the pairs of variables, that is, $E(\mathbf{XY}),$
$E(\mathbf{XZ}),$ $E(\mathbf{YZ}).$ It is well known that for some
sets of angles the joint probability exists, while for other sets
of angles it does not. We can prove that the joint probability doesn't
exist in the following way. We start with the values for the expectations
given by Bell: \begin{eqnarray}
E(\mathbf{XY}) & = & -\frac{\sqrt{3}}{2},\label{eq:marginals-first}\\
E(\mathbf{XZ}) & = & -\frac{\sqrt{3}}{2},\\
E(\mathbf{YZ}) & = & -\frac{1}{2}.\label{eq:marginals-last}\end{eqnarray}
The above expectations correspond to the angles between detectors
set as $\widehat{XY}=30^{\mbox{o}},$ $\widehat{YZ}=30^{\mbox{o}}$,
and $\widehat{XZ}=60^{\mbox{o}}$. It follows from Suppes and Zanotti
(1981) that a joint probability distribution for $\mathbf{X}$, $\mathbf{Y}$,
and \textbf{$\mathbf{Z}$ }as above defined exist if and only if \begin{equation}
-1\leq E(\mathbf{XY})+E(\mathbf{YZ})+E(\mathbf{XZ})\leq1+2\min\left\{ E(\mathbf{XY}),E(\mathbf{YZ}),E(\mathbf{XZ})\right\} .\label{eq:Suppes-Zanotti-inequality}\end{equation}
Clearly, inequalities (\ref{eq:Suppes-Zanotti-inequality}) are violated
for expectations given by (\ref{eq:marginals-first})--(\ref{eq:marginals-last}),
and no joint probability distribution exists. 

What changes with upper probabilities? The system of linear equations
necessary for the existence of a joint distribution becomes a system
of inequalities. This change makes it possible to obtain solutions
to the system, and then upper probabilities that are consistent with
the observed expectations (Suppes and Zanotti, 1991).

\section{Bell-type inequalities for the GHZ state}

As we saw in Section \ref{sec:Upper-Bell-entanglement}, the two-particle
entangled state used by Einstein, Podolski, and Rosen has observables
whose correlations cannot be explained by a joint probability distribution.
In 1989, Greenberger, Horne and Zeilinger (GHZ) concocted a four-system
entangled state that had two new features: the correlations were connected
to path interference and the values of the observables seemed to lead
to mathematical contradictions. The seemingly mathematical contradiction
arose from an assumption of existence of a local hidden-variables
theory that could explain the experimental outcomes predicted by quantum
mechanics. Thus, GHZ proved that quantum mechanics is incompatible
with hidden variables without using inequalities, but instead using
perfect correlations. Their result is known as the GHZ theorem. 

GHZ's argument, as stated by Mermin (1990a), goes as follows. We start
with a three-particle entangled state \begin{equation}
|\psi\rangle=\frac{1}{\sqrt{2}}(|+\rangle_{1}|+\rangle_{2}|-\rangle_{3}+|-\rangle_{1}|-\rangle_{2}|+\rangle_{3}),\end{equation}
where $\hat{\sigma}_{iz}|\pm\rangle_{i}=\pm|\pm\rangle$, and $\hat{\sigma}_{iz}$
is the spin operator in the $\hat{\mathbf{z}}$ direction on the Hilbert
space of the $i$-th particle. This state is an eigenstate of the
following spin operators: \begin{eqnarray}
\hat{\mathbf{A}} & = & \hat{\sigma}_{1x}\hat{\sigma}_{2y}\hat{\sigma}_{3y},\,\,\,\,\hat{\mathbf{B}}=\hat{\sigma}_{1y}\hat{\sigma}_{2x}\hat{\sigma}_{3y},\\
\hat{\mathbf{C}} & = & \hat{\sigma}_{1y}\hat{\sigma}_{2y}\hat{\sigma}_{3x},\,\,\,\,\hat{\mathbf{D}}=\hat{\sigma}_{1x}\hat{\sigma}_{2x}\hat{\sigma}_{3x}.\end{eqnarray}
If we compute the expected values for the above correlations, we obtain
at once that $E(\hat{\mathbf{A}})=E(\hat{\mathbf{B}})=E(\hat{\mathbf{C}})=1$
and $E(\hat{\mathbf{D}})=-1.$ That these correlations present a problem
can be seen from the following theorem (a simplified version of the
theorem found in Suppes, de Barros, and Oas (1998)). 
\begin{description}
\item [{{Theorem~1}}] Let $\mathbf{A},$ $\mathbf{B},$ and $\mathbf{C}$
be three $\pm1$ random variables and let \\
 \textbf{(i)} $E(\mathbf{A})=E(\mathbf{B})=E(\mathbf{C})=1,$ \\
 \textbf{(ii)} $E(\mathbf{ABC})=-1.$ \\
Then \textbf{(i)} and \textbf{(ii)} imply a contradiction. 
\end{description}
\noindent \emph{Proof:} By definition \begin{equation}
E(a)=P(a)-P(\overline{a}),\end{equation}
 where we use the notation $P(a)=P\left(\mathbf{A}=1\right)$ and
$P(\overline{a})=P\left(\mathbf{A}=-1\right)$. Since $0\leq P(a),$
$P(\overline{a})\leq1$, it follows at once from (i) that \begin{equation}
P(a)=1,\label{(1)}\end{equation}
and \begin{equation}
P(\overline{a})=0.\label{eq:}\end{equation}
Similarly, from (ii) and (iii), \begin{equation}
P(b)=P(c)=1.\label{(2)}\end{equation}
\begin{equation}
P(\overline{b})=P(\overline{c})=0.\label{(2)}\end{equation}
 Using again the definition of expectation and the inequalities $P(\overline{a}bc)\leq P(\overline{a})=0,$
etc., we have \begin{equation}
\begin{array}{lll}
E(\mathbf{ABC}) & = & P(abc)+P(\overline{ab}c)+P(a\overline{bc})+P(\overline{a}b\overline{c})\\
 &  & -[P(\overline{a}bc)+P(a\overline{b}c)+P(ab\overline{c})+P(\overline{a}\overline{b}\overline{c})]\\
 & = & 1,\end{array}\label{(3)}\end{equation}
from (\ref{(1)}) and (\ref{(2)}), since all but the first term on
the right are 0. Thus, by conservation of probability, $P(abc)=1$
and the last line follows. But (\ref{(3)}) contradicts (ii). $\diamondsuit$

Of course, the above theorem assumes the existence of an underlying
joint probability distribution. The relationship between the above
theorem and the existence of hidden variables can be illustrated by
the following. Let us now suppose that the value of the spin for each
particle is dictated by a hidden variable $\lambda$, and let us call
this value $s_{ij}(\lambda),$ where $i=1...3$ and $j=x,y.$ Because
spin measurements on each particle can be separated by a space-like
interval, we have that \begin{eqnarray}
E(\hat{\mathbf{A}}\hat{\mathbf{B}}\hat{\mathbf{C}}) & = & (s_{1x}s_{2y}s_{3y})(s_{1y}s_{2x}s_{3y})(s_{1y}s_{2y}s_{3x})\label{Contradiction1}\\
 & = & s_{1x}s_{2x}s_{3x}(s_{1y}^{2}s_{2y}^{2}s_{3y}^{2}).\end{eqnarray}
 Since the $s_{ij}(\lambda)$ can only be $1$ or $-1,$ we obtain\begin{equation}
E(\hat{\mathbf{A}}\hat{\mathbf{B}}\hat{\mathbf{C}})=s_{1x}s_{2x}s_{3x}=E(\hat{\mathbf{D}}).\label{Contradiction2}\end{equation}
But (\ref{Contradiction1}) implies that $E(\hat{\mathbf{A}}\hat{\mathbf{B}}\hat{\mathbf{C}})=1$,
whereas (\ref{Contradiction2}) implies $E(\hat{\mathbf{A}}\hat{\mathbf{B}}\hat{\mathbf{C}})=E(\hat{\mathbf{D}})=-1.$
It should be evident from the above derivation that we could avoid
contradictions if we allow the value of \emph{$\lambda$} to depend
on the experimental setup, i.e., if we allow $\lambda$ to be a contextual
hidden variable. In other words, what the GHZ theorem proves is that
non-contextual hidden variables cannot reproduce quantum mechanical
predictions. 

One of the striking characteristics of GHZ's example is that a contradiction
between quantum mechanics and a hidden-variable theory comes from
probability one (or zero) events. This, however, leads to a problem.
How can we experimentally verify predictions based on correlation-one
events given that experimentally we cannot obtain perfectly correlated
events? This problem was also present in Bell's original paper, where
he didn't consider experimental errors. To {}``avoid Bell's experimentally
unrealistic restrictions'', Clauser, Horne, Shimony and Holt (1969)
derived a new set of inequalities that would take into account imperfections
in the measurement process. However, Bell's inequalities are quite
different from the GHZ case, where it is \emph{necessary} to have
experimentally unrealistic perfect correlations. 

It is important to note that if we could measure all the random variables
simultaneously, we would have a joint distribution. The existence
of a joint probability distribution is a necessary and sufficient
condition for the existence of a hidden variable (Suppes and Zanotti
(1991)). Hence, if the quantum mechanical GHZ correlations are obtained,
then no hidden variable exists. However, this abstract version of
the GHZ theorem still involves probability-one statements. On the
other hand, the correlations present in the GHZ state are so strong
that even if we allow for experimental errors, the non-existence of
a joint distribution, or, equivalently (as shown by Suppes and Zanotti
(1991)) the non-existence of a hidden variable, can still be verified,
as we now proceed to show. We follow de Barros and Suppes (2001).
We start by defining the $\pm1$-valued random variables $\mathbf{X}_{1}$,
$\mathbf{X}_{2}$, $\mathbf{X}_{3}$, $\mathbf{Y}_{1}$, $\mathbf{Y}_{2}$
and $\mathbf{Y}_{3}$ corresponding to the outcomes of spin measurements.
The random variable representing the outcomes of $\hat{\sigma}_{1x}$
is $\mathbf{X}_{1}$, $\hat{\sigma}_{2x}$ is $\mathbf{X}_{2}$, $\hat{\sigma}_{1y}$
is $\mathbf{Y}_{1}$, and so on. Before we derive the inequalities,
we note that if we could measure all the random variables $\mathbf{X}_{1}$,
$\mathbf{X}_{2}$, $\mathbf{X}_{3}$, $\mathbf{Y}_{1}$, $\mathbf{Y}_{2}$
and $\mathbf{Y}_{3}$ simultaneously, we would have a joint probability
distribution. The existence of a joint probability distribution is
a necessary and sufficient condition for the existence of a noncontextual
hidden variable (Suppes and Zanotti, 1991). Hence, if the quantum
mechanical GHZ correlations are obtained, then no such hidden variable
exists. However, Theorem 1 still involves probability-one statements.
On the other hand, the quantum mechanical correlations present in
the GHZ state are so strong that even if we allow for experimental
errors, the non existence of a joint distribution can still be verified,
as we show in the following theorem, which, as we said above, extends
the results in de Barros and Suppes (2000). 
\begin{description}
\item [{Theorem~2}] Let $\mathbf{X}_{i}$ and $\mathbf{Y}_{i}$, $1\leq i\leq3$,
be six $\pm1$ random variables. Then, there exists a joint probability
distribution for all six random variables if and only if the following
inequalities are satisfied:

$-2\leq E(\mathbf{X}_{1}\mathbf{Y}_{2}\mathbf{Y}_{3})+E(\mathbf{Y}_{1}\mathbf{X}_{2}\mathbf{Y}_{3})+E(\mathbf{Y}_{1}\mathbf{Y}_{2}\mathbf{X}_{3})-E(\mathbf{X}_{1}\mathbf{X}_{2}\mathbf{X}_{3})\leq2,$

$-2\leq-E(\mathbf{X}_{1}\mathbf{Y}_{2}\mathbf{Y}_{3})+E(\mathbf{Y}_{1}\mathbf{X}_{2}\mathbf{Y}_{3})+E(\mathbf{Y}_{1}\mathbf{Y}_{2}\mathbf{X}_{3})+E(\mathbf{X}_{1}\mathbf{X}_{2}\mathbf{X}_{3})\leq2,$

$-2\leq E(\mathbf{X}_{1}\mathbf{Y}_{2}\mathbf{Y}_{3})-E(\mathbf{Y}_{1}\mathbf{X}_{2}\mathbf{Y}_{3})+E(\mathbf{Y}_{1}\mathbf{Y}_{2}\mathbf{X}_{3})+E(\mathbf{X}_{1}\mathbf{X}_{2}\mathbf{X}_{3})\leq2,$

$-2\leq E(\mathbf{X}_{1}\mathbf{Y}_{2}\mathbf{Y}_{3})+E(\mathbf{Y}_{1}\mathbf{X}_{2}\mathbf{Y}_{3})-E(\mathbf{Y}_{1}\mathbf{Y}_{2}\mathbf{X}_{3})+E(\mathbf{X}_{1}\mathbf{X}_{2}\mathbf{X}_{3})\leq2.$

\end{description}
\emph{Proof:} The argument is similar to the one found in de Barros
and Suppes (2000). To simplify, we use a notation where $x_{1}y_{2}y_{3}$
means $\mathbf{X}_{1}\mathbf{Y}_{2}\mathbf{Y}_{3}=1$, $\overline{x_{1}y_{2}y_{3}}$
means $\mathbf{X}_{1}\mathbf{Y}_{2}\mathbf{Y}_{3}=-1$. We prove first
that the existence of a joint probability distribution implies the
four inequalities. Then, we have by an elementary probability computation
that \begin{eqnarray*}
P(x_{1}y_{2}y_{3}) & = & P(x_{1}y_{2}y_{3},y_{1}x_{2}y_{3},y_{1}y_{2}x_{3})+P(x_{1}y_{2}y_{3},\overline{y_{1}x_{2}y_{3}},y_{1}y_{2}x_{3})\\
 &  & +P(x_{1}y_{2}y_{3},y_{1}x_{2}y_{3},\overline{y_{1}y_{2}x_{3}})+P(x_{1}y_{2}y_{3},\overline{y_{1}x_{2}y_{3}},\overline{y_{1}y_{2}x_{3}})\end{eqnarray*}
 and \begin{eqnarray*}
P(\overline{x_{1}y_{2}y_{3}}) & = & P(\overline{x_{1}y_{2}y_{3}},y_{1}x_{2}y_{3},y_{1}y_{2}x_{3})+P(\overline{x_{1}y_{2}y_{3}},\overline{y_{1}x_{2}y_{3}},y_{1}y_{2}x_{3})\\
 &  & +P(\overline{x_{1}y_{2}y_{3}},y_{1}x_{2}y_{3},\overline{y_{1}y_{2}x_{3}})+P(\overline{x_{1}y_{2}y_{3}},\overline{y_{1}x_{2}y_{3}},\overline{y_{1}y_{2}x_{3}}),\end{eqnarray*}
 with similar equations for $\mathbf{Y}_{1}\mathbf{X}_{2}\mathbf{Y}_{3}$
and $\mathbf{Y}_{1}\mathbf{Y}_{2}\mathbf{X}_{3}$. But \[
\mathbf{X}_{1}\mathbf{X}_{2}\mathbf{X}_{3}=(\mathbf{X}_{1}\mathbf{Y}_{2}\mathbf{Y}_{3})(\mathbf{Y}_{1}\mathbf{X}_{2}\mathbf{Y}_{3})(\mathbf{Y}_{1}\mathbf{Y}_{2}\mathbf{X}_{3}),\]
 and so we have that \begin{eqnarray*}
P(x_{1}x_{2}x_{3}) & = & P(x_{1}y_{2}y_{3},y_{1}x_{2}y_{3},y_{1}y_{2}x_{3})+P(\overline{x_{1}y_{2}y_{3}},\overline{y_{1}x_{2}y_{3}},y_{1}y_{2}x_{3})\\
 &  & +P(x_{1}y_{2}y_{3},\overline{y_{1}x_{2}y_{3}},\overline{y_{1}y_{2}x_{3}})+P(\overline{x_{1}y_{2}y_{3}},y_{1}x_{2}y_{3},\overline{y_{1}y_{2}x_{3}})\end{eqnarray*}
 and \begin{eqnarray*}
P(\overline{x_{1}x_{2}x_{3}}) & = & P(\overline{x_{1}y_{2}y_{3}},\overline{y_{1}x_{2}y_{3}},\overline{y_{1}y_{2}x_{3}})+P(\overline{x_{1}y_{2}y_{3}},y_{1}x_{2}y_{3},y_{1}y_{2}x_{3})\\
 &  & +P(x_{1}y_{2}y_{3},\overline{y_{1}x_{2}y_{3}},y_{1}y_{2}x_{3})+P(x_{1}y_{2}y_{3},y_{1}x_{2}y_{3},\overline{y_{1}y_{2}x_{3}}).\end{eqnarray*}
 A straightforward computation shows that \begin{eqnarray*}
F & = & 2[P(x_{1}y_{2}y_{3},y_{1}x_{2}y_{3},y_{1}y_{2}x_{3})+P(\overline{x_{1}y_{2}y_{3}},y_{1}x_{2}y_{3},y_{1}y_{2}x_{3})\\
 &  & +P(x_{1}y_{2}y_{3},\overline{y_{1}x_{2}y_{3}},y_{1}y_{2}x_{3})+P(x_{1}y_{2}y_{3},y_{1}x_{2}y_{3},\overline{y_{1}y_{2}x_{3}})]\\
 &  & -2[P(\overline{x_{1}y_{2}y_{3}},\overline{y_{1}x_{2}y_{3}},\overline{y_{1}y_{2}x_{3}})+P(\overline{x_{1}y_{2}y_{3}},\overline{y_{1}x_{2}y_{3}},y_{1}y_{2}x_{3})\\
 &  & +P(\overline{x_{1}y_{2}y_{3}},y_{1}x_{2}y_{3},\overline{y_{1}y_{2}x_{3}})+P(x_{1}y_{2}y_{3},\overline{y_{1}x_{2}y_{3}},\overline{y_{1}y_{2}x_{3})}],\end{eqnarray*}
 where $F$ is defined by \[
F=E(\mathbf{X}_{1}\mathbf{Y}_{2}\mathbf{Y}_{3})+E(\mathbf{Y}_{1}\mathbf{X}_{2}\mathbf{Y}_{3})+E(\mathbf{Y}_{1}\mathbf{Y}_{2}\mathbf{X}_{3})-E(\mathbf{X}_{1}\mathbf{X}_{2}\mathbf{X}_{3}).\]
 Since all probabilities are non-negative and sum to $\leq1,$ we
infer the first inequality at once. The derivation of the other inequalities
is similar.

Now for the sufficiency part. First, we assume the symmetric case
where \begin{equation}
E(\mathbf{X}_{1}\mathbf{Y}_{2}\mathbf{Y}_{3})=E(\mathbf{Y}_{1}\mathbf{X}_{2}\mathbf{Y}_{3})=E(\mathbf{Y}_{1}\mathbf{Y}_{2}\mathbf{X}_{3})=2p-1,\label{14a}\end{equation}
 and \begin{equation}
E(\mathbf{X}_{1}\mathbf{X}_{2}\mathbf{X}_{3})=-(2p-1).\label{14b}\end{equation}
 Then, the first inequality yields \begin{equation}
\frac{1}{4}\leq p\leq\frac{3}{4},\label{inequality2other}\end{equation}
 while the other ones yield \begin{equation}
0\leq p\leq1.\label{inequality2}\end{equation}
 Since $\mathbf{X}_{i}$ and $\mathbf{Y}_{i}$ are $\pm1$ random
variables, $p$ has to belong to the interval $[0,1],$ and inequality
(\ref{inequality2}) doesn't add anything new. We will prove the existence
of a joint probability distribution for this symmetric case by showing
that, given any $p,$ $\frac{1}{4}\leq p\leq\frac{3}{4},$ we can
assign values to the atoms that have the proper marginal distributions. 

The probability space for $\mathbf{X}_{i}$ and $\mathbf{Y}_{i}$
has 64 atoms. It is difficult to handle a problem of this size, so
we will assume some further symmetries to reduce the problem. First,
we introduce the following notation: if a group of symbols is between
square brackets, all the possible permutations of the bar symbol is
considered. For example, $a_{5}=P([\bar{x}_{1}x_{2}x_{3}]y_{1}y_{2}y_{3})$
means that $P(\bar{x}_{1}x_{2}x_{3}y_{1}y_{2}y_{3})=a_{5}$, $P(x_{1}\bar{x}_{2}x_{3}y_{1}y_{2}y_{3})=a_{5}$,
and $P(x_{1}x_{2}\bar{x}_{3}y_{1}y_{2}y_{3})=a_{5}$. Then, the number
of independent values for the probabilities of atoms in the problem
is reduced to the following $16$: $a_{1}=P(x_{1}x_{2}x_{3}y_{1}y_{2}y_{3})$,
$a_{2}=P(x_{1}x_{2}x_{3}\bar{y}_{1}\bar{y}_{2}\bar{y}_{3})$, $a_{3}=P(x_{1}x_{2}x_{3}[\bar{y}_{1}y_{2}y_{3}])$,
$a_{4}=P(x_{1}x_{2}x_{3}[\bar{y}_{1}\bar{y}_{2}y_{3}])$, $a_{5}=P([\bar{x}_{1}x_{2}x_{3}]y_{1}y_{2}y_{3})$,
$a_{6}=P([\bar{x}_{1}x_{2}x_{3}]\bar{y}_{1}\bar{y}_{2}\bar{y}_{3})$,
$a_{7}=P([\bar{x}_{1}x_{2}x_{3}][\bar{y}_{1}y_{2}y_{3}])$, $a_{8}=P([\bar{x}_{1}x_{2}x_{3}][\bar{y}_{1}\bar{y}_{2}y_{3}])$,
$a_{9}=P([\bar{x}_{1}\bar{x}_{2}x_{3}][\bar{y}_{1}y_{2}y_{3}])$,
$a_{10}=P([\bar{x}_{1}\bar{x}_{2}x_{3}][\bar{y}_{1}\bar{y}_{2}y_{3}])$,
$a_{11}=P([\bar{x}_{1}\bar{x}_{2}x_{3}]y_{1}y_{2}y_{3})$, $a_{12}=P([\bar{x}_{1}\bar{x}_{2}x_{3}]\bar{y}_{1}\bar{y}_{2}\bar{y}_{3})$,
$a_{13}=P(\bar{x}_{1}\bar{x}_{2}\bar{x}_{3}[\bar{y}_{1}y_{2}y_{3}])$,
$a_{14}=P(\bar{x}_{1}\bar{x}_{2}\bar{x}_{3}[\bar{y}_{1}\bar{y}_{2}y_{3}])$,
$a_{15}=P(\bar{x}_{1}\bar{x}_{2}\bar{x}_{3}y_{1}y_{2}y_{3})$, $a_{16}=P(\bar{x}_{1}\bar{x}_{2}\bar{x}_{3}\bar{y}_{1}\bar{y}_{2}\bar{y}_{3})$.

These new added symmetries reduce the problem from $64$ to $16$
variables. The atoms have to satisfy various sets of equations. The
first set comes just from the requirement that $E(\mathbf{X}_{i})=E(\mathbf{Y}_{i})=0,$
for $i=1,2,3,$ but two of the six equations are redundant, and so
we are left with the following four. \begin{eqnarray}
a_{1}+a_{2}+3a_{3}+3a_{4}+a_{5}+a_{6}+3a_{7}+3a_{8}-3a_{9}\nonumber \\
-3a_{10}-a_{11}-a_{12}-3a_{13}-3a_{14}-a_{15}-a_{16} & = & 0,\label{SetEqn1}\end{eqnarray}
\begin{eqnarray}
a_{1}-a_{2}+a_{3}-a_{4}+3a_{5}-3a_{6}+3a_{7}-3a_{8}+3a_{9}\nonumber \\
-3a_{10}+3a_{11}-3a_{12}+a_{13}-a_{14}+a_{15}-a_{16} & = & 0,\label{SetEqn2}\end{eqnarray}
\begin{eqnarray}
a_{1}-a_{2}+a_{3}-a_{4}+3a_{5}-3a_{6}+3a_{7}-3a_{8}+3a_{9}\nonumber \\
-3a_{10}+3a_{11}-3a_{12}-a_{13}+a_{14}+a_{15}-a_{16} & = & 0,\label{SetEqn3}\end{eqnarray}
\begin{eqnarray}
a_{1}+a_{2}+3a_{3}+3a_{4}-a_{5}-a_{6}-3a_{7}-3a_{8}+3a_{9}\nonumber \\
+3a_{10}+a_{11}+a_{12}-3a_{13}-3a_{14}-a_{15}-a_{16} & = & 0,\label{SetEqn4}\end{eqnarray}
where (\ref{SetEqn1}) comes from $E(\mathbf{X}_{1})=0,$ (\ref{SetEqn2})
from $E(\mathbf{X}_{2})=0,$ (\ref{SetEqn3}) from $E(\mathbf{Y}_{1})=0,$
and (\ref{SetEqn4}) from $E(\mathbf{Y}_{2})=0.$ The triple expectations
also imply \begin{eqnarray}
a_{1}+a_{2}+3a_{3}+3a_{4}-3a_{5}-3a_{6}-9a_{7}-9a_{8}+9a_{9}\nonumber \\
+9a_{10}+3a_{11}+3a_{12}-3a_{13}-3a_{14}-a_{15}-a_{16} & = & -2p+1,\end{eqnarray}
\begin{eqnarray}
a_{1}+a_{2}-a_{4}-a_{4}+a_{5}+a_{6}-a_{7}-a_{8}+a_{9}\nonumber \\
+a_{10}-a_{11}-a_{12}+a_{13}+a_{14}-a_{15}-a_{16} & = & 2p-1,\end{eqnarray}
and \begin{eqnarray}
a_{1}-a_{2}-3a_{3}+3a_{4}+3a_{5}-3a_{6}-9a_{7}+9a_{8}-9a_{9}\nonumber \\
+9a_{10}+3a_{11}-3a_{12}-3a_{13}+3a_{14}+a_{15}-a_{16} & = & 2p-1.\end{eqnarray}
Finally, the probabilities of all atoms have to sum to one, yielding
the last equation\begin{eqnarray}
a_{1}+a_{2}+3a_{3}+3a_{4}+3a_{5}+3a_{6}+9a_{7}+9a_{8}+9a_{9}\nonumber \\
+9a_{10}+3a_{11}+3a_{12}+3a_{13}+3a_{14}+a_{15}+a_{16} & = & 1.\label{SetEqn9}\end{eqnarray}

Even with the symmetries reducing the problem to 16 variables, we
still have an infinite number of solutions that satisfy equations
(\ref{SetEqn1})--(\ref{SetEqn9}). Since it is very hard to exhibit
a general solution for (\ref{SetEqn1})--(\ref{SetEqn9}) and the
constraints $0\leq a_{i}\leq1,$ $i=1\ldots16$, we will just show
that particular solutions exist for an arbitrary $p$ satisfying the
inequality (\ref{inequality2other}). To do so, we will divide the
problem into two parts: one where we will exhibit an explicit solution
for the atoms $a_{1},\ldots,a_{16}$ that form a proper probability
distribution for $p\in[\frac{1}{4},\frac{1}{2}]$, and another explicit
solution for $p\in[\frac{1}{2},\frac{3}{4}]$. 

It is easy to verify that, given an arbitrary $p$ in $[\frac{1}{4},\frac{1}{2}]$,
the following set of values constitute a solution of equations (\ref{SetEqn1})--(\ref{SetEqn9}):
$a_{1}=0,$ $a_{2}=-\frac{1}{2}+2p,$ $a_{3}=\frac{1}{4}-\frac{1}{2}p,$
$a_{4}=0,$ $a_{5}=0,$ $a_{6}=0,$ $a_{7}=0,$ $a_{8}=0,$ $a_{9}=0,$
$a_{10}=0,$ $a_{11}=0,$ $a_{12}=\frac{1}{4}-\frac{1}{2}p,$ $a_{13}=0,$
$a_{14}=0$, $a_{15}=p,$ $a_{16}=0$. For $p$ in $[\frac{1}{2},\frac{3}{4}]$
the following set of values constitute a solution of equations (\ref{SetEqn1})--(\ref{SetEqn9}):
$a_{1}=-\frac{1}{8}+\frac{1}{2}p,$ $a_{2}=0,$ $a_{3}=\frac{3}{8}-\frac{1}{2}p,$
$a_{4}=0,$ $a_{5}=-\frac{5}{24}+\frac{1}{3}p$, $a_{6}=-\frac{1}{24}+\frac{1}{6}p,$
$a_{7}=0,$ $a_{8}=0,$ $a_{9}=0,$ $a_{10}=0,$ $a_{11}=0,$ $a_{12}=0,$
$a_{13}=0,$ $a_{14}=\frac{1}{8}$, $a_{15}=\frac{3}{8}-\frac{1}{2}p,$
$a_{16}=0$. So, for $p$ satisfying the inequality $\frac{1}{4}\leq p\leq\frac{3}{4}$
we can always construct a probability distribution for the atoms consistent
with the marginals, and this concludes the proof. $\diamondsuit$ 

We note that the form of the inequalities of Theorem 2 is actually
that of the Clauser et al. (1969) for the Bell case, when the Bell
binary correlations are replaced by the GHZ triple correlations. The
inequalities from Theorem 2 immediately yield the following. 
\begin{description}
\item [{Corollary}] Let $\mathbf{X}_{i}$ and $\mathbf{Y}_{i}$, $1\leq i\leq3$,
be six $\pm1$ random variables, and let

\begin{description}
\item [{(i)}] $E(\mathbf{X}_{1}\mathbf{Y}_{2}\mathbf{Y}_{3})=E(\mathbf{Y}_{1}\mathbf{X}_{2}\mathbf{Y}_{3})=E(\mathbf{Y}_{1}\mathbf{Y}_{2}\mathbf{X}_{3})=1-\varepsilon,$
\item [{(ii)}] $E(\mathbf{X}_{1}\mathbf{X}_{2}\mathbf{X}_{3})=-1+\varepsilon$, 
\end{description}
\noindent $\varepsilon\in[0,1].$ Then there cannot exist a joint
probability distribution of $\mathbf{X}_{i}$ and $\mathbf{Y}_{i}$,
$1\leq i\leq3$, satisfying (i) and (ii) if $\varepsilon<\frac{1}{2}$.

\end{description}
\noindent \emph{Proof.} If a joint probability exists, then \[
-2\leq E(\mathbf{X}_{1}\mathbf{Y}_{2}\mathbf{Y}_{3})+E(\mathbf{Y}_{1}\mathbf{X}_{2}\mathbf{Y}_{3})+E(\mathbf{Y}_{1}\mathbf{Y}_{2}\mathbf{X}_{3})-E(\mathbf{X}_{1}\mathbf{X}_{2}\mathbf{X}_{3})\leq2.\]
But \[
E(\mathbf{X}_{1}\mathbf{Y}_{2}\mathbf{Y}_{3})+E(\mathbf{Y}_{1}\mathbf{X}_{2}\mathbf{Y}_{3})+E(\mathbf{Y}_{1}\mathbf{Y}_{2}\mathbf{X}_{3})-E(\mathbf{X}_{1}\mathbf{X}_{2}\mathbf{X}_{3})=4-4\varepsilon,\]
and the inequality is satisfied only if $\varepsilon\geq\frac{1}{2}$.
Hence, if $\varepsilon<\frac{1}{2}$ no joint probability exists.
$\diamondsuit$

In the Corollary, $\varepsilon$ may represent, for instance, a deviation
from the predicted quantum mechanical correlations due to experimental
errors. So, we see that to prove the nonexistence of a joint probability
distribution for the GHZ experiment, we do not need to have perfect
measurements and $1$ or $-1$ correlations. In fact, from the above
inequalities, it should be clear that any experiment that satisfies
the strong symmetry of the Corollary and obtains a correlation for
the triples stronger than $0.5$ (and $-0.5$ for one of them) cannot
have a joint probability distribution. 

It is worth mentioning at this point that the inequalities derived
in Theorem 2 have a completely different origin than do Bell's inequalities.
The inequalities of Theorem 2 are not satisfied by a particular model,
but they just accommodate the theoretical conditions in GHZ to possible
experimental deviations. Also, Theorem 2 does not rely on any {}``enhancement''
hypothesis to reach its conclusion. Thus, with this reformulation
of the GHZ theorem it is possible to use strong, yet imperfect, experimental
correlations to prove that a hidden-variable theory is incompatible
with the experimental results.

\section{Upper probabilities and the GHZ state}

We now analyze the existence of upper probabilities for the GHZ state.
The following theorem states our main result. 
\begin{description}
\item [{{Theorem~3}}] Let $\mathbf{A},$ $\mathbf{B},$ and $\mathbf{C}$
be three $\pm1$ random variables and let \\
 \textbf{(i)} $E^{*}(\mathbf{A})=E(\mathbf{A})=1,$ \\
 \textbf{(ii)} $E^{*}(\mathbf{B})=E(\mathbf{B})=1,$\\
 \textbf{(iii) $E^{*}(\mathbf{C})=E(\mathbf{C})=1,$}\\
 \textbf{(iv)} $E^{*}(\mathbf{ABC})=E(\mathbf{ABC})=-1.$ \\
 Then, there exists an upper joint probability distribution that
is compatible with expectations (i)--(iv). 
\end{description}
\emph{Proof:} We prove the theorem by explicitly providing an upper
joint probability distribution. Let \begin{equation}
p^{*}\left(abc\right)=p^{*}\left(\overline{a}\overline{b}\overline{c}\right)=1\label{eq:joint-1}\end{equation}
and \begin{equation}
p^{*}\left(\overline{a}bc\right)=p^{*}\left(a\overline{b}c\right)=p^{*}\left(ab\overline{c}\right)=p^{*}\left(a\overline{b}\overline{c}\right)=p^{*}\left(\overline{a}b\overline{c}\right)=p^{*}\left(\overline{a}\overline{b}c\right)=0.\label{eq:joint-0}\end{equation}
Since $E^{*}\left(\mathbf{A}\right)=E^{*}\left(\mathbf{B}\right)=E^{*}\left(\mathbf{C}\right)=1$,
it follows that \begin{equation}
p^{*}\left(a\right)=p^{*}\left(b\right)=p^{*}\left(c\right)=1\label{eq:1-upper-prob}\end{equation}
and \begin{equation}
p^{*}\left(\overline{a}\right)=p^{*}\left(\overline{b}\right)=p^{*}\left(\overline{c}\right)=0.\label{eq:0-upper-prob}\end{equation}
 Next, let us consider the events\begin{equation}
a=abc\cup a\overline{b}c\cup ab\overline{c}\cup a\overline{b}\overline{c},\label{eq:a-joint-atoms}\end{equation}
\begin{equation}
\overline{a}=\overline{a}bc\cup\overline{a}\overline{b}c\cup\overline{a}b\overline{c}\cup\overline{a}\overline{b}\overline{c},\label{eq:abar-joint-atoms}\end{equation}
and similarly for $b$, $\overline{b}$, $c$, and $\overline{c}$.
From (\ref{eq:a-joint-atoms}) and the subadditive properties of the
upper distributions must hold, and we have \[
p^{*}\left(a\right)\leq p^{*}\left(abc\right)+p^{*}\left(a\overline{b}c\right)+p^{*}\left(ab\overline{c}\right)+p^{*}\left(a\overline{b}\overline{c}\right).\]
 Using (\ref{eq:joint-1}), (\ref{eq:joint-0}), and (\ref{eq:1-upper-prob}),
the above inequality becomes\[
1\leq1+0+0+0,\]
consistent with the joint. For (\ref{eq:abar-joint-atoms}), the subadditive
properties requires \[
p^{*}\left(\overline{a}\right)\leq p^{*}\left(\overline{a}bc\right)+p^{*}\left(\overline{a}\overline{b}c\right)+p^{*}\left(\overline{a}b\overline{c}\right)+p^{*}\left(\overline{a}\overline{b}\overline{c}\right).\]
 Using (\ref{eq:joint-1}), (\ref{eq:joint-0}), and (\ref{eq:0-upper-prob}),
the above inequality becomes\[
0\leq0+0+0+1,\]
also consistent. Similar computations follow for $b$, $\overline{b}$,
$c$, and $\overline{c}$. 

Going back to the expectation, we are given \[
E\left(\mathbf{ABC}\right)=-1,\]
or \begin{equation}
E^{*}\left(\mathbf{ABC}\right)=1\cdot p^{*}\left(abc\cup\overline{a}\overline{b}c\cup\overline{a}b\overline{c}\cup a\overline{b}\overline{c}\right)+\left(-1\right)\cdot p^{*}\left(\overline{a}bc\cup a\overline{b}c\cup ab\overline{c}\cup\overline{a}\overline{b}\overline{c}\right).\label{eq:joint-expectation}\end{equation}
From (\ref{eq:joint-expectation}) \begin{equation}
p^{*}\left(abc\cup\overline{a}\overline{b}c\cup\overline{a}b\overline{c}\cup a\overline{b}\overline{c}\right)=0,\label{eq:joint-atom-union-0}\end{equation}
and from (\ref{eq:joint-1}), (\ref{eq:joint-0}), (\ref{eq:joint-atom-union-0}),
and the subadditive properties, \begin{eqnarray*}
p^{*}\left(abc\cup\overline{a}\overline{b}c\cup\overline{a}b\overline{c}\cup a\overline{b}\overline{c}\right) & \leq & p^{*}\left(abc\right)+p^{*}\left(\overline{a}\overline{b}c\right)+p^{*}\left(\overline{a}b\overline{c}\right)+p^{*}\left(a\overline{b}\overline{c}\right)\\
0 & = & 1+0+0+0.\end{eqnarray*}
Also, from (\ref{eq:joint-expectation}), \begin{equation}
p^{*}\left(\overline{a}bc\cup a\overline{b}c\cup ab\overline{c}\cup\overline{a}\overline{b}\overline{c}\right)=1,\label{eq:joint-atom-union-1}\end{equation}
and from (\ref{eq:joint-1}), (\ref{eq:joint-0}), (\ref{eq:joint-atom-union-1}),
and the subadditive properties,\begin{eqnarray*}
p^{*}\left(\overline{a}bc\cup a\overline{b}c\cup ab\overline{c}\cup\overline{a}\overline{b}\overline{c}\right) & \leq & p^{*}\left(\overline{a}bc\right)+p^{*}\left(a\overline{b}c\right)+p^{*}\left(ab\overline{c}\right)+p^{*}\left(\overline{a}\overline{b}\overline{c}\right)\\
1 & = & 0+0+0+1.\end{eqnarray*}
Thus, we complete the check of all probabilities necessary for consistency.
The remaining events can easily be assigned upper probabilities that
satisfy the axioms of upper probabilities. $\diamondsuit$

\section{Conclusions}

To apply the upper probabilities to the GHZ theorem, we gave a probabilistic
random variable version of it. We then showed that, if we use upper
probabilities, some of the lemmas used to derive GHZ do not hold anymore,
and hence the inconsistencies cannot be proved to exist from the upper
probabilities. But upper probabilities are a natural way to deal with
contextual problems in statistics. For example, in pools in social
sciences upper probabilities can be applied, as the results of the
responses depend on the context of the pool. This contextuality in
the GHZ derivation was shown in the previous section. Therefore, generic
contextual hidden variable theories do not result in any logical contradictions,
as often claimed, since the mathematical contradictions derived in
GHZ do not appear when we weaken the requirements for probabilities,
allowing them to be upper probabilities instead.

\section{References}

ASPECT, A., DALIBARD, J., and ROGER, G. {}``Experimental test of
Bell's inequalities using time-varying analyzers''. Physical Review
Letters, 49, pp. 1804--1807, 1982. \\
BELL, J. S. Speakable and Unspeakable in Quantum Mechanics. Cambridge:
Cambridge University Press, 1987. \\
BOUWMEESTER, D., PAN, J-W., DANIELL, M., WEINGURTER, H., and ZEILINGER,
A. {}``Observation of three-photon Greenberger-Horne-Zeilinger entanglement''.
Physical Review Letters, 82, pp. 1345--1349, 1999. \\
BRICMONT, J., DURR, D. GALAVOTTI, M. C., GHIRARDI, G., PETRUCCIONE,
F, and ZANGHI, N. Chance in Physics: Foundations and Perspectives.
Berlin: Springer, 2001.\\
CLAUSER, J. F., HORNE, M. A., SHIMONY, A., and HOLT, R. A. {}``Proposed
experiment to test local hidden variable theories''. Physical Review
Letters 23, pp. 880--884, 1969.\\
DE BARROS, J. A and SUPPES, P. {}``Inequalities for dealing with
detector inefficiencies in GHZ-type experiments''. Physical Review
Letters, 84, pp. 793--797, 2000. \\
DE BARROS, J. A and SUPPES, P. {}``Probabilistic results for six
detectors in a three-particle GHZ experiment''. In: Bricmont et al.
(eds.) (2001), pp. 213--223.\\
EINSTEIN, A., PODOLSKI, B., and ROSEN, N. {}``Can the quantum
mechanical description of physical reality be considered complete?''.
Physical Review 47, pp. 777--780, 1935. \\
GARG, A. and MERMIN, N. D. {}``Correlation inequalities and hidden
variables,'' Physical Review Letters, 49, pp. 1220--1223, 1982. \\
GREENBERGER, D. M., HORNE, M., and ZEILLINGER, A. {}``Going beyond
Bell's theorem''. In: Kafatos (ed.) (1989), pp. 73--76.\\
KAFATOS, M. Bell's Theorem, Quantum Theory, and Conceptions of
the Universe. Dordrecht: Kluwer Academic Publishing, 1989. \\
MERMIN, N. D. {}``Quantum mysteries revisited''. American Journal
of Physics, 58, pp. 731--734, 1990a. \\
MERMIN, N. D. {}``Extreme quantum entanglement in a superposition
of macroscopically distinct states''. Physical Review Letters, 65,
pp. 1838--1840, 1990b. \\
PERES, A. Quantum Theory: Concepts and Methods. Dordrecht: Kluwer
Academic Publishing, 1995. \\
PRATESI, R. and RONCHI, L. Waves, Information and Foundations of
Physics. Bologna: Italian Physics Society, 1998. \\
SUPPES, P., DE BARROS, J. A, and OAS, G. {}``A collection of probabilistic
hidden-variable theorems and counterexamples''. In: R. Pratesi and
L. Ronchi (eds.) (1998), pp. 267--291.\\
SUPPES P. and ZANOTTI, M. {}``When are probabilistic explanations
possible?''. Synthese, 48, pp. 191--199, 1981. \\
SUPPES P. and ZANOTTI, M. {}``Existence of hidden variables having
only upper probabilities''. Foundations of Physics, 21, pp. 1479--1499,
1991. 
\end{document}